\documentclass[a4paper,11pt]{article}
\pdfoutput=1 % if your are submitting a pdflatex (i.e. if you have
\usepackage{jcappub} % for details on the use of the package, please
\usepackage[T1]{fontenc} % if needed

\title{\boldmath Scalar Field as a Bose--Einstein Condensate? \footnote{Submitted to JCAP}}

\author[a,b]{El\'ias Castellanos,}
\author[a]{Celia Escamilla-Rivera,}
\author[c]{Alfredo Mac\'ias}
\author[d]{and Dar\'{\i}o N\'u\~nez}

\affiliation[a]{Mesoamerican Centre for Theoretical Physics (ICTP regional headquarters in Central America, the Caribbean and Mexico), Universidad Aut\'onoma de Chiapas, Carretera Zapata Km. 4, Real del Bosque (Ter\'an), 29040, Tuxtla Guti\'errez, Chiapas, M\'exico.}
\affiliation[b]{Departamento de F\'{\i}sica, Centro de Investigaci\'on y Estudios Avanzados
del IPN,\\A. P. 14--740, 07000, M\'exico, D.F., M\'exico.}
\affiliation[c]{Departamento de F\'{\i}sica, Universidad Aut\'onoma Metropolitana--Iztapalapa,\\A.P. 55-534, Mexico D.F. 09340, M\'exico.}
\affiliation[d]{Instituto de Ciencias Nucleares, Universidad Nacional
Aut\'onoma de M\'exico, Circuito Exterior C.U., A.P. 70-543,
M\'exico D.F. 04510, M\'exico.}
\emailAdd{ecastellanos@mctp.mx}
\emailAdd{cescamilla@mctp.mx}
\emailAdd{amac@xanum.uam.mx}
\emailAdd{nunez@nucleares.unam.mx}

\abstract{We discuss the analogy between a classical scalar field with a self--interacting potential, in a
curved spacetime described by a quasi--bounded state, and a trapped Bose--Einstein condensate. In this context, we 
compare the Klein--Gordon equation with the Gross--Pitaevskii 
equation. Moreover, the introduction of a curved background  spacetime 
endows, in a natural way, an equivalence to the Gross--Pitaevskii equation with an explicit confinement 
potential. The curvature also induces a position dependent self--interaction parameter. We exploit this 
analogy by means of the Thomas--Fermi approximation,
commonly used to describe the Bose--Einstein condensate, in order to analyze the quasi bound scalar field distribution 
surrounding a black hole.}

\begin{document}
\maketitle
\flushbottom

\section{Introduction}
\label{sec:intro}
\indent
Since half a century ago, when the Jordan--Brans--Dicke scalar--tensor theory of gravity was proposed 
\cite{Jordan:1961vv, BD61}, scalar fields 
have experienced a long and controversial life. Nowadays, they appear in the formulation of many phenomena 
in gravitational theories.
On the one hand, a scalar field is always present in the context of Dirac's large number hypothesis, and also in 
all higher--dimensional unified field theories; they appear as dilatons in string theory and as inflatons or dark matter 
in cosmo\-lo\-gy \cite{bransSF97}. Nevertheless, they have remained until now as exotic matter. It was only in the
last year that the Higgs boson was detected \cite{Higgs:1964ia, Brout:1969hh, Abreu:2011zz}, a very important fact in
the development of the scalar field theory.  

It has also been found that there exist fundamental relations between particle physics, cosmology and condensed matter 
\cite{unruh81}.
Different condensed matter systems -- such as acoustics in flowing fluids, light in moving dielectrics or quasi particles
in moving superfluids -- can be shown to reproduce some aspects of General Relativity and cosmology 
\cite{volovik2003}. They can be conceived as laboratory toy models that make  
some features of quantum field theory on curved spacetime experimentally accessible \cite{Schtzhold2008}. 

In this context, one of the most peculiar phenomena in physics, discovered in the last
century, is Bose--Einstein condensation \cite{Bose,Einstein}, which is a macroscopic quantum phenomenon that 
was first discovered theoretically
by Bose \cite{Bose} and Einstein \cite{Einstein} in the $1920's$, where it
was applied through the new concept of Bose statistics to a non--interacting gas of identical atoms
which were at thermal equilibrium and trapped in a box. It was predicted that,
at sufficiently low temperatures, the particles would accumulate in the lowest
quantum state in the box and would merge into a giant superatom. Locked
together, moving as one, this condensate of atoms would become a new state
of matter, different from solid, liquid or gas. Thus, large numbers of bosons can collapse into the same 
quantum state to form
a condensate, while two fermions cannot be in the same quantum state -- they obey the Pauli principle 
\cite{pathria}. Bose--Einstein condensation is only possible for massive bosonic particles.
The particle density at the center
of a Bose--Einstein  condensate is typically of the order
$10^{13}$--$10^{15} cm^{-3}$ \cite{dalfobo-giorgini-pita-string99}, not that far from the density
of molecules in air at room temperature and atmospheric pressure, which is
of the order $10^{19} cm^{-3}$, or the density of atoms in liquids and solids, which is
about $10^{22} cm^{-3}$ \cite{Vandevenne2005, pethick2002}. It was a great achievement
when this theoretical idea was finally realized in the laboratory, 75 years 
later \cite{Anderson:1995gf, Davis:1995pg, Bradley:1995zz}.

The Gross--Pitaevskii equation \cite{gross61,pita61} is usually applied to investigate the
physical properties of Bose--Einstein condensates in trapped ultra--cold atoms with a
temperature, $T$, of about 100 $nK$. This equation
was derived independently by Gross \cite{gross61} and Pitaevskii \cite{pita61} in 1961.
Its validity is based on the conditions that, for a diluted gas, the s--wave scattering length must be much smaller than 
the average distance between atoms, and that the
number of atoms in the condensate must be much larger than one. It can be used at very 
low temperature (including absolute zero) to
explore the macroscopic behavior of the system, characterized by variations of the order parameter over distances
larger than the mean distance between atoms \cite{dalfobo-giorgini-pita-string99,zoller00}.

It is interesting to note that, after several approximations and assumptions, it can be shown that the 
Klein--Gordon equation governing the dynamics of a classical scalar field can be reduced to the Schr\"odinger equation, 
from which the Gross--Pitaevskii equation that governs the dynamics of a Bose--Einstein condensate follows. 
It is also remarkable that, as mentioned above, we face a classical relativistic scalar field (with second order 
time derivatives), and a quantum Newtonian BEC (with first order time derivatives), with 
different concepts of time \cite{maque06}. In this
way, the Klein--Gordon equation contains the Gross--Pitaevskii one. However, the physical meaning being described 
changes quantitatively and qualitatively between these ideas. 

Even though the Schr\"odinger equation can be derived from the Klein--Gordon equation, they
describe very different physical phenomena. There are, however, macroscopic and quantum systems 
which, under certain circumstances, are 
described by the same formal equation, with the same mathematical structure. This is the case between a scalar 
field distribution surrounding a black hole, the so called
quasi--bound states \cite{Barranco11}, and the Bose--Einstein condensate described by the Gross--Pitaevskii
equation. We remark that in the former case, it is the space--time curvature, together with the scalar field features, that
form an effective potential which contains the scalar field distribution.

At least until now, it has not been proper to call the quasi bound distributions `cosmological condensates', as long as such
distributions are described by a classical field without any reference to particles or quantum states. Even though 
the quasi bound distributions satisfy a very similar equation to the one satisfied by the stationary Bose--Einstein 
condensate, they have a very different origin, describe very different phenomena and, as mentioned
above, are conceptually different. The similarity between the equations which describes each 
case is, however, worthy of deeper research, see for example \cite{Khlopov:1985jw}. In fact, the scalar field description of 
dark matter and its possible relation with Bose--Einstein condensates is yet to be 
fully understood \cite{urena}.

In this paper we will use this similarity to investigate the possible implications in 
determining the dynamics of a scalar field,
using the techniques applied in solving the dynamics of a Bose--Einstein condensate. 
The mentioned similarity will work in static 
and stationary situations where the role played by the time coordinate is not relevant. We will see how it is 
possible to express, under certain circumstances, the Klein--Gordon equation
as a Gross--Pitaevskii--like equation. As we mentioned above, the presence of a gravitational background provides, 
in a natural way, a trapping potential in the effective Gross--Pitaevskii equation. Once the analogy is established, we will 
apply the usual techniques of atomic physics to the description of the quasi bound state.

We consider the pathological one--dimensional case, where the Thomas--Fermi approximation breaks down, but we 
are able to obtain analytical solutions for the different descriptions, so that we can, in principle, compare them.

Finally, we must mention that the scalar field collapse can form stable compact objects: an oscillaton in the case of a real 
scalar field, and a boson star if the collapsing scalar field is complex; see for instance \cite{BS}. These configurations 
could also be compared with Bose-Einstein condensates, and we could try to find analogies.  However, in this work 
we are stressing the mathematical similarity between the final equation describing quasi-stationary scalar distributions in a curved background, and the Gross-Pitaevskii one. An analogy between the Bose-Einstein condensate and the physical 
compact object, such as the boson star, could also be an interesting line worth pursuing, but it is beyond the scope 
of the present paper and will be discussed in future works.

The outline of the paper is as follows: In Sec. II, the flat space analogy between a classical scalar field 
and a Bose--Einstein condensate is described. In Sec. III, a curved space analogy is considered. In Sec. IV the Thomas--Fermi approximation 
is applied to the scalar field equation in a gravitational background. In Sec. V, the conclusions and outlook are presented.

%--------------------------------------------------------------------------------------------------------------------------------------------
%--------------------------------------------------------------------------------------------------------------------------------------------
\section{Flat space analogy}

Let us first consider the flat spacetime case, \emph{i.e.,} without a gravitational background. We will find an 
exact solution of the Gross--Pitaevskii equation, obtained from the Klein--Gordon 
equation, for the case of a scalar field trapped in a one--dimensional box. It is worthwhile to mention 
that the solution has the form of the 
one obtained for the case of a Bose--Einstein condensate trapped in a one--dimensional box. This solution is 
known as the static soliton. 

%--------------------------------------------------------------------------------------------------------------------------------------------
%--------------------------------------------------------------------------------------------------------------------------------------------

\subsection{One--dimensional scalar field trapped in a box: Static soliton}

First, we reduce the Klein--Gordon equation for a classical scalar field to a Gross--Pitaevskii--like equation for a Bose--Einstein condensate in one spatial dimension.
Following \cite{rusa}, we analyze the problem of a scalar field trapped in a box.

It is worthwhile to note that in the one dimensional flat case, the Gross--Pitaevskii--like equation does not have a confinement 
potential other than the boundary walls of the box.

Consider a complex scalar field, $\Phi(t,x)$, satisfying the following Klein--Gordon equation
\begin{equation}
-\frac1{c^2}\,\ddot{\Phi} + \Phi^{\prime\prime} - \frac{d\,V_{1d}}{d\,\Phi^*}=0\, , \label{eq:KG1d}
\end{equation}
where the prime denotes a derivative with respect to $x$, and the dot a derivative with respect to $c\,t$.
We restrict ourselves to the case of harmonic time dependence of a scalar field, which is coupled to a 
scalar self--interacting potential
\begin{eqnarray}
\Phi(t,x)&=&e^{i\,\omega\,t}\,\chi(x), \label{eq:Phi1d} \\
V_{1d}&=&\frac{\sigma^2}{2}\,\Phi\,\Phi^* +\frac{\lambda}{4}\,(\Phi\,\Phi^*)^2, \label{eq:V1d}\\
&=&\frac{\sigma^2}{2}\,\chi^2 +\frac{\lambda}{4}\,\chi^4. \label{eq:V1da}
\end{eqnarray}
Additionally, the condition $(d\,V_{1d}/d\Phi^*)=e^{i\,\omega\,t}\,(d\,V_{1d}/d\chi)$ reduces the Klein--Gordon 
equation to the following Gross--Pitaevskii--like equation; see 
Eq.~(\ref{3dGross--Pitaevskii equation0}) below (the generalization to three dimensions is straightforward, see for 
instance Refs. \cite{Elias9})
\begin{equation}
\label{chi0}
\chi^{\prime \prime} - \left[\left(\sigma^2 - \frac{\omega^2}{c^2} \right)\,\chi + \lambda\,\chi^3\right]=0,
\end{equation}
which can be integrated directly leading to
\begin{equation}
\frac12\,{\chi'}^2 - \frac{\lambda}{4}\,\left(\chi^2 + \frac{\sigma^2 - \frac{\omega^2}{c^2} }{\lambda}\right)^2=C_1
\label{papajohn}\, ,
\end{equation}
where we added a term $(-[\sigma^2 - (\omega^2/c^2)]^2/4\,\lambda)$ to
both sides of the equation and defined a new integration cons\-tant $C_1$.

Eq. (\ref{papajohn}) can be solved directly, leading to a JacobiSN solution, which should be restricted to the 
case of vanishing integration constants. Thus, by imposing analogous conditions for the relation between the chemical potential $\mu$ and the parameter that describes the interaction $U_{0}=4 \pi \hbar^{2}a/m$, with $a$ being the s--wave scattering length as for the usual Gross--Pitaevskii equation, \emph{i.e.},  $\mu=U_0|\psi(x)|^2=U_0\,n$, and $n$ the corresponding particle density, we obtain the following condition for our system 
\begin{equation}
\label{muefe}
\mu_{\rm eff} \equiv \frac{\omega^2}{c^2} - \sigma^2= \lambda\,|\chi_0|^2\, \equiv \lambda\,\rho_{N},
\end{equation}
where we have defined an effective chemical potential $\mu_{eff}$, together with a new constant $|\chi_0|^2$, which corresponds, as in the usual case, to the wave function far away from the wall, where the kinetic energy term becomes negligible.

The balance between the kinetic term and the interaction energy 
characterized by the coupling constant $\lambda$ in Eq. (\ref{chi0}) allows us to fix a typical distance 
over which the system can \emph{heal}, as in the usual Gross--Pitaevskii equation. In our case this is
\begin{equation}
\label{HEFL}
\xi_{flat}=\frac{1}{\sqrt{\lambda \rho_{N}}}.
\end{equation}
It is interesting to notice that the associated healing length $\xi_{flat}$ is independent of the parameter $\sigma$ due to the 
functional form of Eq.(\ref{chi0}).

Finally, in this scenario, the solution to the Klein--Gordon equation for a classical scalar field with 
a self--interacting potential reduces to following expression 
\begin{equation}
\label{tan-tan}
\chi(x) = |\chi_0| \tanh(x/\sqrt{2}\xi_{\rm flat}),
\end{equation}
which is precisely the kink solution obtained for a Bose--Einstein condensate trapped in a box \cite{pethick2002}.

We see that in this simple case, for a harmonic--like solution, and an infinite barrier potential, we can relate
the Klein--Gordon equation, and the scalar field solution, to the Gross--Pitaevskii equation 
and the order parameter (that is, the solution to the Gross--Pitaevskii equation inside the 
potential). The chemical potential, $\mu_{eff}$
is identified with the subtraction of the mass parameter and the oscillation 
frequency, as seen in Eq.~(\ref{muefe}).

%--------------------------------------------------------------------------------------------------------------------------------------------
%--------------------------------------------------------------------------------------------------------------------------------------------

\section{Curved Space Analogy}

\indent 
In order to continue exploring the analogy between the Klein--Gordon equation and the Gross--Pitaevskii equation, we 
now present the Klein--Gordon equation in a curved spacetime. 

One starts from the Lagrangian density for a complex scalar field
\begin{equation}
{\cal L}=\frac{c^4}{16\,\pi\,G}\,\,\left(\frac12\,\nabla\,\Phi\,\nabla\,\Phi^* - V(\Phi\,\Phi^*)\right), \label{eq:LagPhic}
\end{equation}
where $V(\Phi\,\Phi^*)$ is the scalar field potential. From this Lagrangian one obtains, by 
$\partial\,{\cal L}/\partial\,g^{\mu\,\nu}$, the stress energy tensor:
\begin{eqnarray}
T_{\mu\,\nu}&=&\frac{c^4}{16\,\pi\,G}\,\left[\Phi_\mu\,{\Phi^*}_\nu + \Phi_\nu\,{\Phi^*}_\mu - g_{\mu\,\nu}\left(g^{\alpha\,\beta}\,\Phi_\alpha\,{\Phi^*}_\beta + 2\,V(\Phi\,\Phi^*) \right) \right],
\label{eq:Tmunu}
\end{eqnarray}
which can be used in Einstein's equations, 
and its conservation equation, ${T^\mu}_{\nu\,;\mu}=0$ lead us to the Klein-Gordon equation for the 
complex scalar field.

The classical scalar field can also be considered as a test particle/field, \emph{i.e.}, it only 
feels gravity, while its own gravity is neglected.
Its dynamics is determined by a Klein--Gordon equation in a curved background spacetime. In this work we consider
the scalar field as a test particle/field in a curved background.

In the previous section we displayed, for a box potential as container, an analogy between a classical scalar field 
satisfying a Klein--Gordon equation and the Gross--Pitaevskii equation for a Bose--Einstein condensate. This fact reinforces our 
guess about the existence of an analogy between
classical scalar field configurations with the order parameter associated with the Bose--Einstein condensate theory.
Now we show that this analogy is actually even more remarkable.

Consider a spherically--symmetric--static background spacetime
\begin{equation}
ds^2=-F(r)\,c^2\,d\,t^2 + \frac{dr^2}{F(r)} + r^2\,d\Omega^2, \label{eq:ele}
\end{equation}
with $d\Omega^2=d\theta^2 +\sin^2\theta\,d\varphi^2$, and $c$ the speed of
light in vacuum. By solving the vacuum Einstein field equations including cosmological constant
\begin{equation}
R_{\mu\,\nu}-\frac12\,g_{\mu\,\nu}\,R + \Lambda\,g_{\mu\,\nu}=0\, \label{eineqs},
\end{equation}
one can determine the explicit form of the geometric function, $F$.

We consider the dynamics of a scalar test field, $\Phi$, with a scalar self--interacting potential given by
\begin{equation}
V(\Phi\,\Phi^*)=\frac{\sigma^2}{2}\,\Phi^*\,\Phi +\frac{\lambda}{4}\,[\Phi^*\,\Phi]^2.\label{eq:V}
\end{equation}

That is, the scalar field satisfies a Klein--Gordon equation in the curved spherically symmetric
background spacetime given by Eq.~(\ref{eq:ele}), which reads
\begin{equation}
\big[g^{\mu\,\nu}\,\nabla_\mu\,\nabla_\nu-\left(\sigma^2 + \lambda\,\rho_n\right)\,\big]\Phi=0, \label{eq:KG1}
\end{equation}
where we used the following definition of the number density $\rho_n$
\begin{equation}
\rho_n=\Phi^*\,\Phi. \label{eq:rhon}
\end{equation}

Let us restrict our attention to the monopolar component of the scalar field with harmonic time dependence
\begin{equation}
\Phi=e^{i\,\omega\,t}\,\frac{u(r)}{r}.\label{eq:phiu}
\end{equation}
The Klein--Gordon equation reduces to a non-linear Schr\"odinger--like equation, which is
a kind of Gross--Pitaevskii--like equation
\begin{equation}
\Bigg(-\frac{d^2\,}{d\,{r^*}^2} + V_{\rm eff} + \lambda\,F\,\rho_n\Bigg)\,u=\frac{\omega^2}{c^2}\,u. \label{eq:KGGP}
\end{equation}
Here, we identified the particle/field density $\rho_n=u^2/r^2$, and introduced the $r^*$ coordinate
\begin{equation}
r^*=\int\,\frac{d\,r}{F}. \label{eq:r_es}
\end{equation}
Hence, the effective trapping potential reads
\begin{equation}
V_{\rm eff}=F\,\,\left(\sigma^2 + \frac{F'}{r^*} \right),\label{eq:Veff}
\end{equation}
where now the prime stands for a derivative with respect to $r^*$.

In order to obtain stationary (or quasi stationary) solutions for the scalar field, the curvature of the spacetime
itself should confine the scalar field. Indeed, it is not necessary to introduce ``by hand'' an external potential to
confine the scalar field in the Klein--Gordon equation; the gravitational background can do the work, with some 
background spacetimes able to
confine the scalar field. In what follows we set $r^{*}\rightarrow r$ to simplify the notation.

Eq.~(\ref{eq:KGGP}) is a Gross--Pitaevskii--like equation
\begin{equation}
\label{3dGross--Pitaevskii equation0}
-\frac{\hbar^{2}}{2m} \mathbf{\nabla}^{2}\psi(\bold{r})+V(\bold{r})\psi(\bold{r})
+U_{0}\vert \psi(\bold{r})\vert ^{2}\psi(\bold{r})=\mu\psi(\bold{r}).
\end{equation}
We identify the effective potential $V_{\rm eff}$ of Eq.~(\ref{eq:KGGP}) with the trapping 
potential $V(\bold{r})$
of the Gross--Pitaevskii equation, Eq.(\ref{3dGross--Pitaevskii equation0}). The Klein-Gordon self--interaction 
term is given by $\lambda\,F$, which
includes a geometric coefficient. In our Gross--Pitaevskii--like equation, this is 
identified with $U_0$. The particle density $\rho_n$ is identified
with $\vert \psi(\mathbf{r})\vert ^{2}$; the frequency,  $(\omega/c)^{2}$, together with the mass 
parameter, $\sigma^2$, as we saw in the 1D case, is identified with the chemical 
potential, $\mu$; and the radial dependence of the scalar field $u$ is related to the order parameter, $\psi$.

Notice from Eq.(\ref{eq:KGGP}) that the curvature of the space--time induces also a spatially dependent 
interaction, through the parameter $\lambda F$. This \emph{effective} interaction 
parameter $\lambda F$ (position dependent) could be interpreted as some 
kind of \emph{gravitational} Feshbach resonance induced by the curvature of 
space--time, and could affect, for instance, the stability of the system, as in usual 
condensates \cite{pethick2002}. Clearly, this particular issue deserves deeper analysis,
to be presented elsewhere, due to the fact that it is intimately related with the structure 
of the system and could be tested, in principle, as dark matter in more realistic scenarios.

In the cases where the effective potential holds the scalar field quasi--stationary configuration, the
analogy between the scalar field in a curved background, satisfying Eq.~(\ref{eq:KGGP}), and the order 
parameter describing a Bose-Einstein condensate, satisfying the Gross-Pitaevskii stationary 
equation, Eq.~(\ref{3dGross--Pitaevskii equation0}), is remarkable.

We present relevant examples of such spacetimes in the following subsections.

%--------------------------------------------------------------------------------------------------------------------------------------------
%--------------------------------------------------------------------------------------------------------------------------------------------

\subsubsection{Schwarzschild--de Sitter spacetime}

Let us consider the case of a Schwarzschild black hole within a de Sitter spacetime, for which the metric 
coefficient $F$ of Eq.~(\ref{eq:ele})
has the form
\begin{equation}
F=1 - \frac{2\,M\,G}{c^2\,r} - \frac{\Lambda}{3}\,r^2, \label{eq:FSdS}
\end{equation}
where $M$ is the mass of the black hole and $\Lambda$ is the cosmological constant.
Choosing a mass scale, $M_0$, and a distance scale,
$R_0$, we construct the dimensionless quantity $q=G\,M_0/c^2\,R_0$, and the mass of the black hole
under study is then a factor of the mass scale, $M=n\,M_0$, and the distance is a multiple of the distance scale
$r=x\,R_0$, with $n, x$ as dimensionless constants. Since $\Lambda$ has units of curvature -- that
is, inverse area -- we construct the dimensionless quantity $\nu= \Lambda\,{R_0}^2/3$, so the
metric coefficient $F$ given in Eq.~(\ref{eq:FSdS}) reduces to the following dimensionless form
\begin{equation}
F=1 - 2\,q\,\Bigl(\frac{n}{x}\Bigr) - \nu\,x^2. \label{eq:FSdS1}
\end{equation}

Assume that $\Lambda$ represents the dark energy in our model. Using Planck data \cite{Planck}, 
the definition for the critical density of the universe, $\rho_{\rm critical}=3\,{H_{0\,{\rm Planck}}}^2/(8\,\pi\,G)$ 
and the gravitational constant value $G=4.29\,\times\,10^{-9}\, ({\rm km/s})^2\,{\rm Mpc/M_{\rm Sun}}$ \cite{Nunez13}, we obtain the value $\rho_{\rm critical}=1.26\,\times\,10^{11}\,{\rm M_{\rm Sun}/Mpc^3}$.
According to Planck priors, the ratio ${\rho_\Lambda}/{\rho_{\rm critical}}=68.3\%$, standing for the dark energy sector, is a quantity that we can use to compute the 
corresponding value for the cosmological constant via $\Lambda={8\pi G}{c^{-2}}\rho_\Lambda$. Finally, we obtain the 
value $\Lambda=1.036\,\times\,10^{-7}\pm{4.302\times 10^{-9}}\,{\rm Mpc^{-2}}$. Choosing the
distance scale as $R_0=1 {\rm Mpc}$, the dimensionless metric coefficient takes the value $\nu=3.452\,\times\,10^{-8}\pm{1.434\times 10^{-9}}$.

In the Schwarzschild--de Sitter case, we deal with two kinds of horizons; one 
related to the black hole, and the other associated with $\Lambda$. This 
external horizon, as long as the cosmological constant $\Lambda$ has a definite
value, is fixed, and amounts to $x_{\rm ext}=5386.37$ in the absence of black hole, which gives the size of a Universe 
dominated by the cosmological constant,  $R_{\rm max}=5.38\,\times 10^3\,{\rm Mpc}$. When a black 
hole is present, and if we 
consider $q=1$, this implies that our mass scale is $10^{19}$ solar masses. In this case there exist two horizons, 
and when one considers a more massive black hole,  
the internal horizon grows towards the external one in such a way that, for a value of $n=1036.6$, the
two horizons merge into one, and we have a critical Schwarzschild--de Sitter spacetime.

Now, the effective potential, Eq.~(\ref{eq:Veff}) for this spacetime reads
\begin{equation}
V_{\rm eff_{SdS}}=\frac{1}{{R_0}^2}\,\left(\alpha^2 - 2\,\nu + \frac{2\,q\,n}{x^3} \right)\,
\left(1 - \frac{2\,q\,n}{x} - \nu\,x^2\right),\label{eq:Veff_SdS}
\end{equation}
where we define $\alpha=R_0\,\sigma$. For the scalar potential we use the same distance scale as 
the one used for the
spacetime parameters, so that the parameter $\alpha$ is dimensionless. For $\alpha<\sqrt{2\,\nu}$, the 
asymptotic behavior of
the effective potential is positive, and for $x\equiv 2\,q\,n$ we have a characteristic black hole barrier.
Thus, we expect to have regions where bound states of the scalar distribution could exist. 

\begin{figure}[tbp]
\centering % \begin{center}/\end{center} takes some additional vertical space
%\includegraphics[width=.45\textwidth,trim=0 380 0 200,clip]{img1.pdf}
%\hfill
\includegraphics[width=1.\textwidth,origin=c,angle=0]{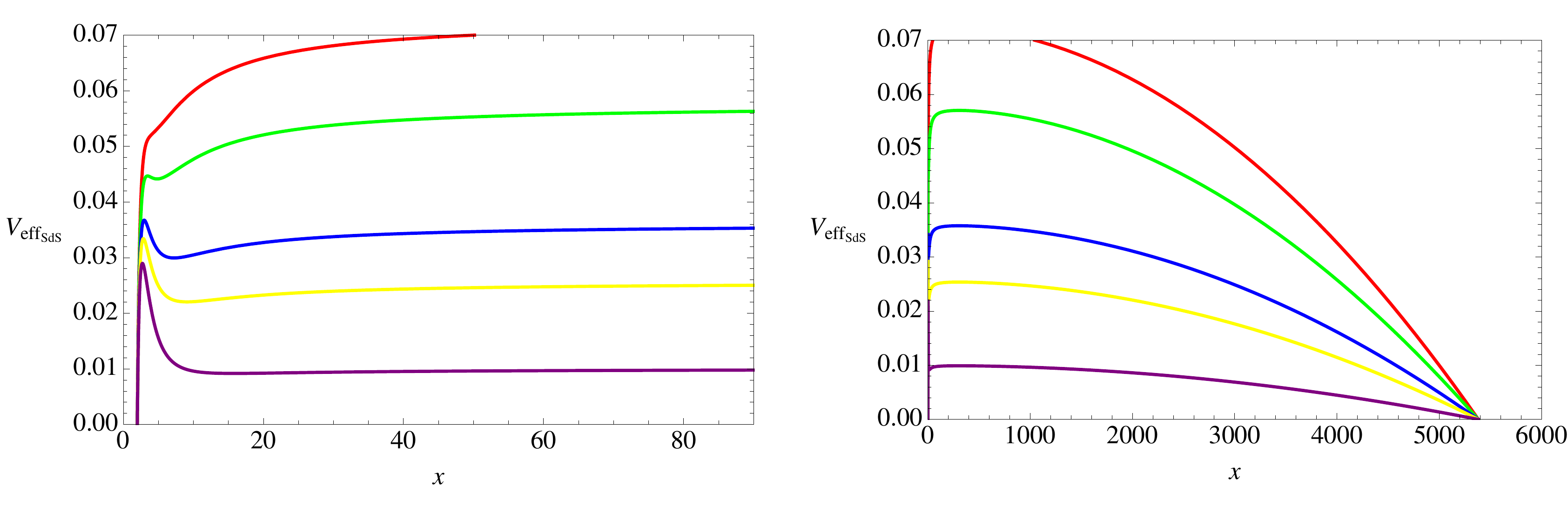}
% "\includegraphics" is very powerful; the graphicx package is already loaded
\caption{\label{fig:Pot_SchwdS2} Effective potential Eq.(\ref{eq:Veff_SdS}), for a large black hole mass
and scalar mass parameter $\alpha$. We see how the potential forms confinement regions depending 
on the values of $\alpha$ and $\Lambda$. Here $q=n=1$ and $\nu=3.452\times 10^{-8}$.  
\textit{Left}: The five curves represent the potential behavior for $\alpha=0.27,0.24, 0.19, 0.16,0.1$, in 
descending order.
For the first value there are no bounded regions and the latter value disappears. \textit{Right}: The corresponding 
behavior for large radii, $10^3\leq x $, in which the potential decreases until it reaches the cosmological horizon 
at $x_{\rm ext}=5386.37$, where it is equal to zero for any value of $\alpha$.}
\end{figure}

We also present a less realistic case, characterized by a huge black hole mass and a very large scalar mass parameter. Despite
its unrealistic features, it is interesting that this limit shows clearly how the effective potential induces the 
formation of bounded regions near the black hole, depending on the value of the scalar mass parameter. For larger radius 
the effective potential grows and then starts to decline, reaching zero value at the cosmological horizon. The results are 
presented in Fig.~(\ref{fig:Pot_SchwdS2}).
In this way, the gravitational field generated by the black hole mass, the cosmological constant,  $\Lambda$,
and the scalar field mass parameter $\alpha$, is endowed with trapped regions for the scalar field.
In the case of a black hole spacetime, \cite{Barranco12} showed that these trapped regions can host 
quasi--stationary distributions of the scalar field, lasting even for cosmological periods.

This fact also strengths the analogy between the scalar field in curved backgrounds and the order parameter describing Bose--Einstein condensates, since there do exist quasi--stationary distributions of the scalar field in a curved background, which
behaves in an analogous way to a Bose--Einstein condensate.

%--------------------------------------------------------------------------------------------------------------------------------------------
%--------------------------------------------------------------------------------------------------------------------------------------------

\subsubsection{Schwarzschild spacetime}

\begin{figure}[ht]
\begin {center}
\includegraphics[height=4.6cm]{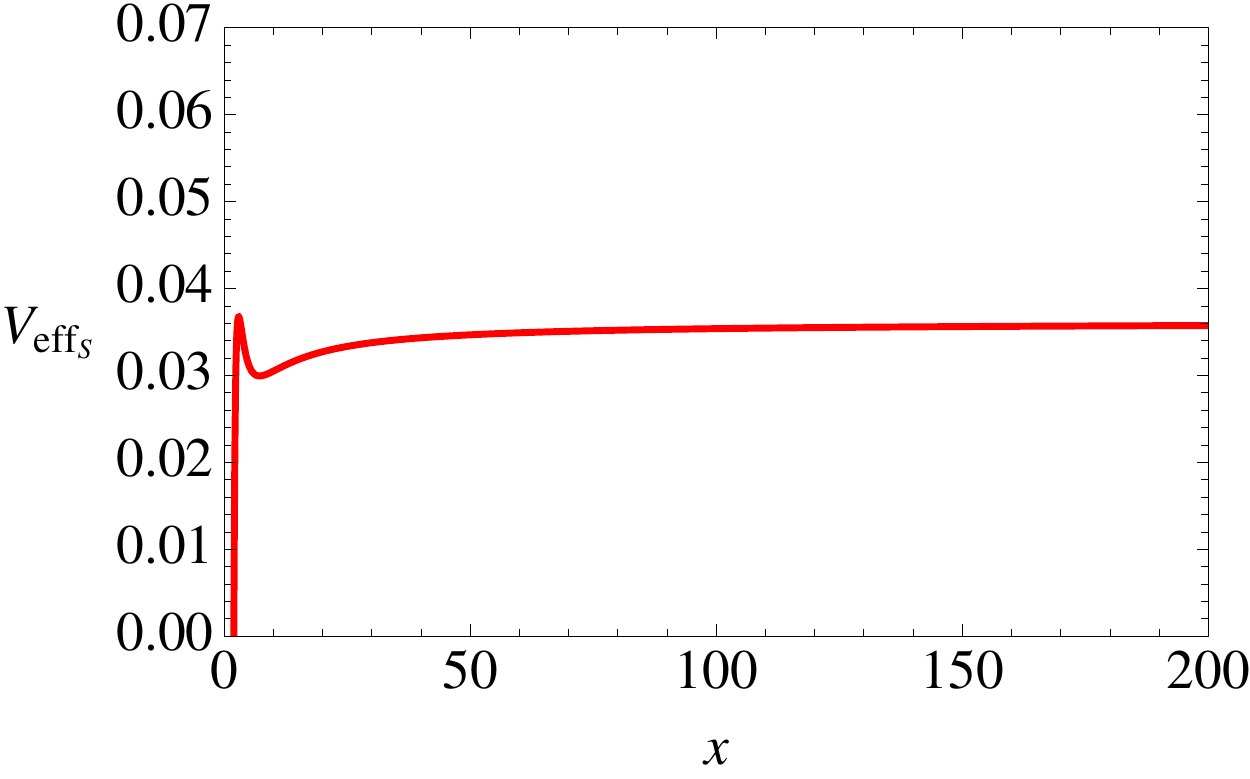}
\caption{Effective potential Eq.(\ref{eq:Veff_S}). We have have taken
${R_0}^2=q=n=1$, and the case $\alpha=0.17$. We observe the confinement region for small values of $x$.}
\label{fig:Pot_Schw}
\end {center}
\end{figure}

This case has been already discussed in detail \cite{Barranco11,Barranco12} within the context of scalar field 
distributions which remain surrounding a black hole for cosmological times. The metric coefficient in the 
line element, Eq.~(\ref{eq:ele}) reduces to the form ($\Lambda=\nu=0$)
\begin{equation}
F=1 - \frac{2\,M\,G}{c^2\,r}=1 - 2q\,\Bigl(\frac{n}{x}\Bigr), \label{eq:FS}
\end{equation}
where now the mass $M_0$ and distance scale $R_0$ are not constrained to be large.
The effective potential
Eq.~(\ref{eq:Veff}) reduces to the following form
\begin{equation}
V_{\rm eff_{S}}=\frac{1}{{R_0}^2}\,\left(\alpha^2 + \frac{2\,q\,n}{x^3} \right)\,\left(1 - \frac{2\,q\,n}{x} \right),\label{eq:Veff_S}
\end{equation}
where the asymptotic value of the effective potential is the square of the scalar mass parameter $\alpha^2$. In this case, the
potential presents confinement regions, as shown in Fig.~(\ref{fig:Pot_Schw}). A detailed discussion has been given
in Refs.\cite{Barranco11,Barranco12}, describing how one does indeed have quasi--stationary scalar field distributions which
accrete towards the black hole, but at such a slow rate that they can last for cosmological times.
%%

We remark that the solutions are quasi--stationary as long as the stationary solutions are forbidden by the black hole
no--hair theorems \cite{macias2012}. Indeed, such scalar field distributions are not completely stationary, and so 
are not excluded by those
theorems; they are not ``hair'', and can last for very long times, depending on the mass of the black black hole and 
the scalar mass parameter. Hence they have been dubbed ``wigs'' \cite{Barranco12}

%--------------------------------------------------------------------------------------------------------------------------------------------
%--------------------------------------------------------------------------------------------------------------------------------------------

\subsubsection{de Sitter spacetime}

The usual cosmological solution to Einstein field equations, Eq.(\ref{eineqs}), proposed originally by de Sitter, reads
\begin{equation}
ds^2=-c^2\,dT^2 +e^{\Lambda\,T}\,\left(d\,R^2 + R^2\,d\,\Omega^2\right)\, , \label{eq:eledSc}
\end{equation}
which can be rewritten as a static--like line element with the form given by Eq.~(\ref{eq:ele}), with
\begin{equation}
F(r)=1 - \frac{\Lambda}{3}\,r^2=1 - \nu\,x^2, \label{eq:FdS}
\end{equation}
so that the effective potential and the coordinate transformation to $r^*$ reduces to the following form
\begin{equation}
V_{\rm eff_{dS}}=\frac{1}{{R_0}^2}\,\left(\alpha^2 - 2\,\nu  \right)\,\left(1- \nu\,x^2\right),\label{eq:Veff_dS}
\end{equation}

\begin{figure}[ht]
\begin {center}
\includegraphics[height=4.6cm]{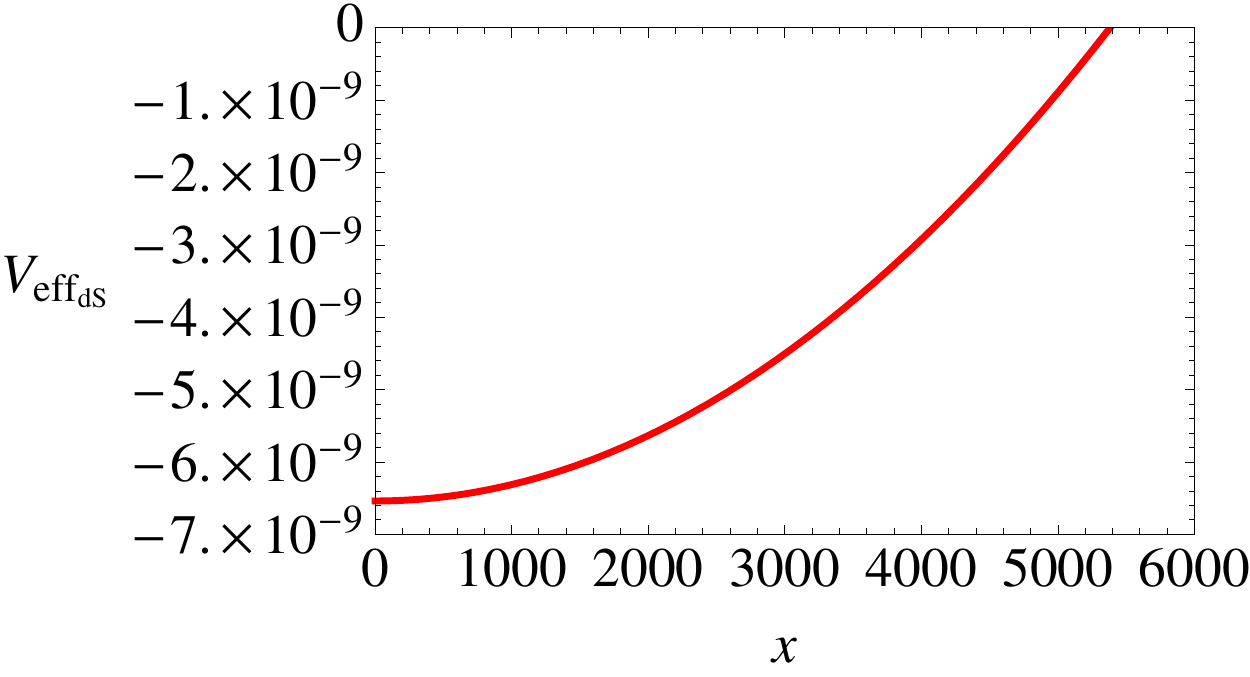}
\caption{Effective potential Eq.(\ref{eq:Veff_dS}). We have have taken
${R_0}^2=q=n=1$ and $\nu=3.452\,\times\,10^{-8}$. The case for  $\alpha=2.5\,\times\,10^{-4}$ shows a 
confinement region where the effective potential takes negative values.}
\label{fig:Pot_dS}
\end {center}
\end{figure}

The asymptotic behavior in this case reads $-\left(\alpha^2 - 2\,\nu  \right)\,\nu\,x^2/{R_0}^2$, and
as long as there is only one extreme, $x=0$, the scalar mass parameter, $\alpha^2$, must be less than the cosmological
one, $\nu$, in order to have a trapped region. This behavior is  shown in Fig.~(\ref{fig:Pot_dS}).

We do not expect to have scalar field distributions in this case, as long as the potential is always negative.

%--------------------------------------------------------------------------------------------------------------------------------------------
%--------------------------------------------------------------------------------------------------------------------------------------------

\section{Thomas--Fermi Approximation for the Scalar Field Equation in Curved Backgrounds}

Following the analogy between the quasi--stationary scalar field distributions in curved background and Bose--Einstein 
condensates described by the Gross--Pitaevskii equation, we are able to explore how the Thomas--Fermi approximation 
can be used to study and, in some sense, to experimentally observe scalar field distributions in curved spacetimes. 
The Thomas--Fermi approximation is very useful for exploring some relevant thermodynamical properties of Bose--Einstein 
condensates in the presence of interactions. In usual condensates, an accurate description of the 
system may be obtained by neglecting the kinetic energy term in the Gross--Pitaevskii equation from the very 
beginning. Such an approximation is valid for sufficiently large clouds, and when the scattering length $a$, which 
describes the interaction among the particles within the system, is much smaller that the mean inter--particle spacing; in 
other words, when the system is diluted enough and contains a large number of atoms. Finally, we must add that the 
Thomas--Fermi approximation fails for trapped condensates near 
the edge of the cloud, due to the divergent behavior of the kinetic energy (\emph{i.e.} the total kinetic energy per unit area 
diverges on the boundary of the system).  

If we assume that the Thomas--Fermi approximation is valid for our system, then we can obtain from 
Eq.~(\ref{eq:KGGP}) the following expression
\begin{equation}
\big(V_{\rm eff} + \lambda\,F\,\rho_n\big)\,u=\frac{\omega^2}{c^2}\,u\, ,
\end{equation}
where we have used the definition given in Eq.~(\ref{eq:rhon}) and, consistently with the analogy between the
classical scalar field and the condensate, we call $\rho_{n}$ the \emph{particle density} which, with the plane wave ansatz,
Eq.~(\ref{eq:phiu}), is given by $\rho_{n}=u^{2}(r)/r^{2}$.

Within the Thomas--Fermi approximation, one transforms a differential equation into an
algebraic one, with a solution
\begin{equation}
\rho_{n}=\frac{\frac{\omega^{2}}{c^{2}}-V_{eff}}{\lambda F}. \label{eq:rhonTF}
\end{equation}
Finding the values where $\rho_n=0$, we obtain the region where the scalar field is contained in this approximation. It is clear 
that there are differences between the solution
obtained within this approximation and the actual solution to the Klein--Gordon equation in curved space-times. 
For example, Eq.~(\ref{eq:rhonTF}) diverges at the horizon with $F=0$, which is an unacceptable behavior for a stable 
scalar field distribution. However, this sort of problem already occurs
in the Thomas--Fermi approximation in usual condensates. The approximation is not valid at the borders. We can
expect, at most, that the Thomas--Fermi approximation describes the scalar field distribution in the regions where
the density has a maximum -- roughly, where the potential has a minimum.
 
There are several definitions of density used in each context and, as we are 
using ideas from different fields, it is important to have clear definitions from each, and to understand how they are related to
one another.

On the one hand, there is the particle density in nuclear physics,  $\rho_{n}$, Eq.~(\ref{eq:rhon}), used in the Thomas--Fermi approximation, Eq.~(\ref{eq:rhonTF}), which is related to the probability density from quantum mechanics. On the other hand, there
is the energy density defined in
the relativistic context: $c^2\,\rho_E=-{T_0}^0$. From Eq.~(\ref{eq:Tmunu}), it is straightforward to show, 
for the space time given by Eq.~(\ref{eq:ele}), that the energy density takes the form
\begin{equation}
\rho_E = \frac{c^2}{16\,\pi\,G}\,\left(\frac{\dot{\Phi}\,\dot{\Phi^*}}{c^2\,F} + F\,{\Phi'}\,{\Phi^*}' 
+ \sigma^2\,\Phi\,{\Phi^*} + \frac{\lambda}{2}\,(\Phi\,{\Phi^*} )^2 \right). \label{eq:rhoE1}
\end{equation}

This is the expression for the density in General Relativity for static and spherically symmetric spacetimes. One can 
see that it includes terms
involving the particle density from nuclear physics, ${\Phi}\,{\Phi^*}$, and also spatial and temporal
derivatives 
which, according to the theory, also determine the energy. Indeed, 
using the plane wave ansatz and the normalization for the scalar function given in Eq. (\ref{eq:phiu}), we obtain the relationship between the energy density and the particle density,
Eq.~ (\ref{eq:rhon})
\begin{equation}
\rho_E = \frac{c^2}{16\,\pi\,G}\,\left[\left(\sigma^2 + \frac{\omega^2}{c^2\,F} \right)\,\rho_n +
\frac{\lambda}{2}\,{\rho_n}^2 + F\,\frac{{{\rho_n}'}^2}{4\,\rho_n}\right], \label{eq:rhoE2}
\end{equation}
The above can define a {\it mass density}, $\rho_{\rm mass}$ (with units of density), related to the
particle density,
\begin{equation}
\rho_{\rm mass} = \frac{c^2\,\sigma^2}{16\,\pi\,G}\,\rho_n. \label{eq:rhomn}
\end{equation}
To obtain the order of magnitude of this mass density, we can consider an ultra-light scalar field
with a mass of around $m_\phi\,c^2 \equiv 10^{-24}\,{\rm eV}$. For $\hbar\,\sigma/c = m_\phi$, we find that
the corresponding parameter $\sigma$ for this ultra-light scalar mass is $5.06\,10^{-18}\,{\rm m}^{-1}$, and $c^2\,\sigma^2/16\,\pi\,G = 6.86\,10^{-13} {\rm grs}/{\rm cm}^3$. Notice that as long as the scalar
field is dimensionless, the particle density is defined as $\rho_n=\Phi\,\Phi^*=u^2/r^2$.

When studying the validity of the Thomas--Fermi approximation in the Klein--Gordon equation in curved spacetimes 
in this way, the first question is: How 
is the number density related to the actual energy density?
The energy density reduces to the mass density under certain conditions. For instance, consider a weak gravitational 
field region, $F\equiv 1$, where the mass density has small gradients and the
self--interaction term is negligible compared with the first order term. In this case,
\begin{equation}
\rho_E \sim \left(1 + \frac{\omega^2}{\sigma^2\,c^2}\right)\,\rho_{\rm mass}. \label{eq:rhoEm}
\end{equation}
which is a relation that could, in principle, be probed, once one has the solution to the Klein-Gordon equation.

Following this line of thought, we also define an expected size and
a possible number of particles for the scalar field in a curved background.

We consider the extremes of the density, defined for those values of the distance, $x_{i}$ and $x_{f}$,
where the frequency, $\omega$, intersects the effective potential. Then, we define the {\it number of particles}
as the integral of the density between those extremes by using Eq.~(\ref{eq:rhomn})
\begin{equation}
\label{NP}
N=\frac{c^2\,\sigma^2}{4\,G}\,\int_{x_i}^{x_f} \rho_{\rm mass} \,x^2\,d\,x.
\end{equation}

Let us now apply these concepts to the concrete spacetimes we already presented.

%--------------------------------------------------------------------------------------------------------------------------------------------
%--------------------------------------------------------------------------------------------------------------------------------------------
\subsection{Schwarzschild--de Sitter spacetime}

Within the Thomas--Fermi approximation, a direct application of the
particle density definition leads to the following expression for the particle density of the scalar field
\begin{figure}[ht]
\begin {center}
\includegraphics[height=5.4cm]{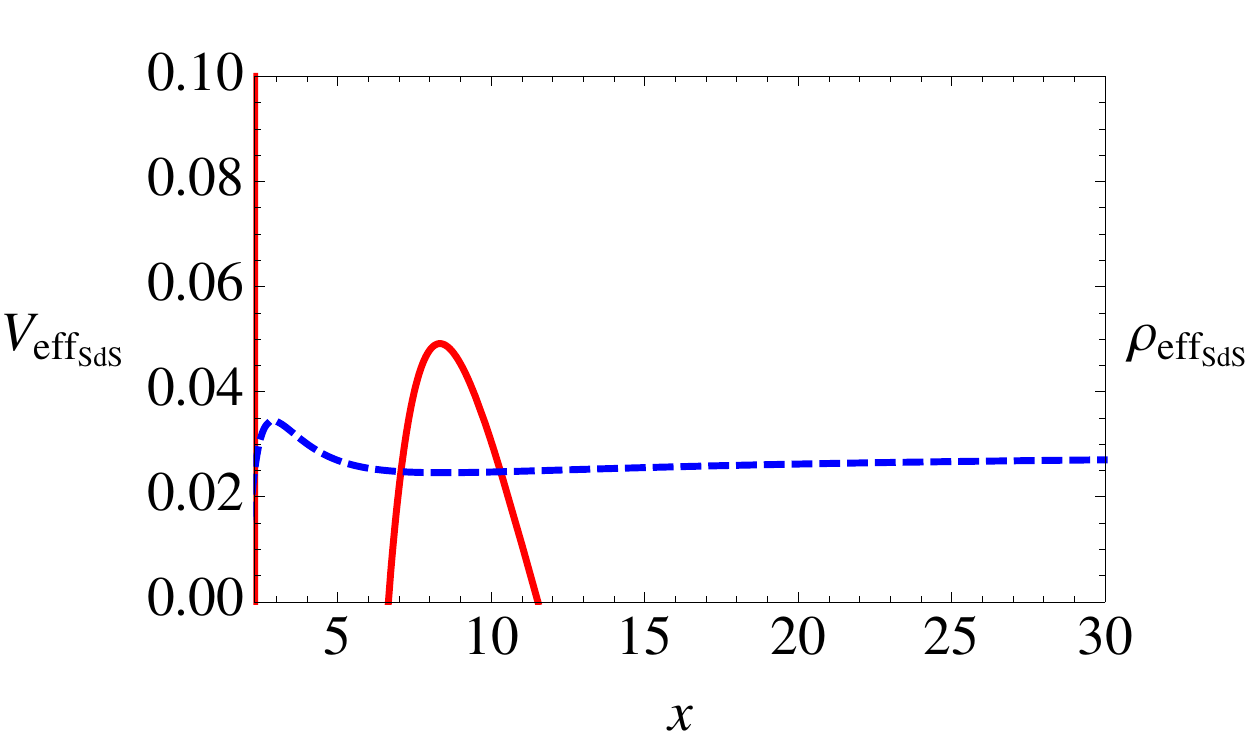}
\caption{Density distribution Eq.(\ref{eq:rhoTF_SdS}) (solid line) and the effective potential Eq.(\ref{eq:Veff_SdS}) (dashed line). The values considered for this case are
$q=n=1, \alpha=0.17, \omega=0.158, \lambda=0.01$.}
\label{fig:Pot_rho_SdS}
\end {center}
\end{figure}

\begin{eqnarray}
{\rho_{n}}_{\rm Sch-deSitt}&=&\frac{\frac{\omega^2}{c^2} - \frac{1}{{R_0}^2}\,
\left(\alpha^2 - 2\,\nu + \frac{2\,q\,n}{x^3} \right)\,\left(1 - \frac{2\,q\,n}{x} - \nu\,x^2\right)\, }{\lambda\,\left(1 - \frac{2\,q\,n}{x} - \nu\,x^2\right)} .\label{eq:rhoTF_SdS}
\end{eqnarray}

From the effective potential plot Fig.~(\ref{fig:Pot_SchwdS2}), we can choose a value for the frequency
which intersects the potential, thus determining the extrema of the density. Then, choosing the value of the
parameter $\lambda$, one obtains a sketch of the mass density distribution. This procedure results in the behavior
shown in Fig.~(\ref{fig:Pot_rho_SdS}).

The cosmological constant modifies the shape of the effective potential in a manner that is noticeable in the external region.
It is also worth noting that it can be interpreted as modifying the value of the mass parameter of the scalar field,
making it ``lighter'', as seen by comparing the effective potential in this case, Eq.~(\ref{eq:Veff_SdS}), and
the one in Schwarzschild, Eq.~(\ref{eq:Veff_S}).

%--------------------------------------------------------------------------------------------------------------------------------------------
%--------------------------------------------------------------------------------------------------------------------------------------------
\subsection{Schwarzschild spacetime}

\begin{figure}[ht]
\begin {center}
\includegraphics[height=5.4cm]{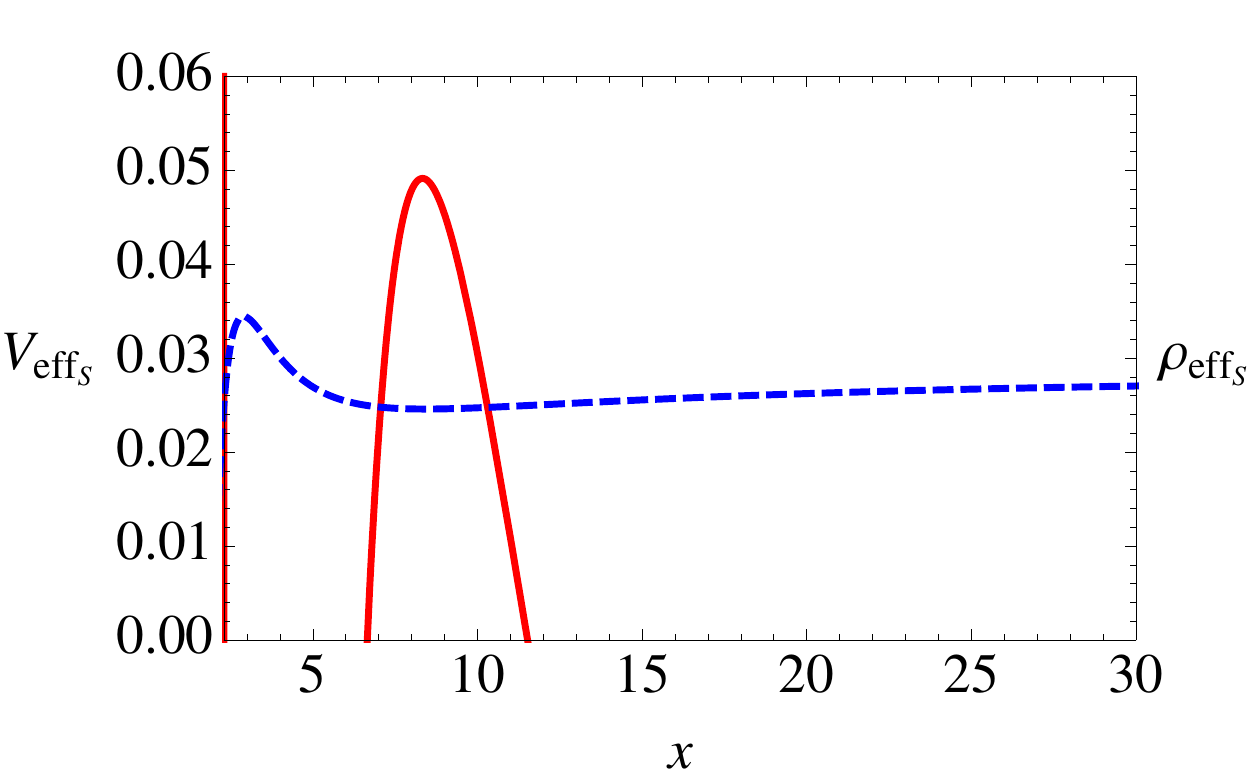}
\caption{Density distribution Eq.(\ref{eq:rhoTF_S}) (solid line) and the effective potential Eq.(\ref{eq:Veff_S}) (dashed line). The values considered for this case are
$q=n=1, \alpha=0.17, \omega=0.158, \lambda=0.01$.}
\label{fig:Pot_rho_S}
\end {center}
\end{figure}

For this case the function $F$ takes the form given by Eq.(\ref{eq:Veff_S}). Then,
the solution reads
\begin{equation}
{\rho_{n}}_{\rm Sch}=\frac{\frac{\omega^2}{c^2} - \frac{1}{{R_0}^2}\,\left(\alpha^2 + \frac{2\,q\,n}{x^3} \right)\,
\left(1 - \frac{2\,q\,n}{x} \right)}{\lambda\,\left(1 - \frac{2\,q\,n}{x}\right)}.\label{eq:rhoTF_S}
\end{equation}
A sketch of the density Eq.(\ref{eq:rhoTF_S}) is presented in Fig.(\ref{fig:Pot_rho_S}).

It is remarkable how straightforward it is to derive the density distribution in this approximation. According to it, the scalar 
field forms a shell surrounding the black hole, with a maximum at a few Schwarzschild radii. One expects that the actual 
solution has this form, but that it goes to zero at the horizon and fades smoothly in the external region. 

In this way, we expect that the Thomas-Fermi approximation describes the shape, and 
perhaps even the position, near the maximum of the density distribution of a quasi stationary 
scalar field surrounding a black hole.
A work comparing the density derived from the actual numerical solution of the 
Gross--Pitaevskii equation with the Thomas-Fermi approximation shows that the 
approximation is indeed excellent in all regions except the extrema \cite{Claus13}. We do not 
expect such excellent accordance, but we think it certainly hints at the 
actual behaviour of the scalar field distribution in curved spacetimes.

Let us add that the trapping potential in usual condensates induces a density peak at the center 
of the trap, \emph{i.e.} near to $r \sim 0$, and in this scenario, the effects caused by interactions, in a dilute 
system, are expected to be significant \cite{dalfobo-giorgini-pita-string99}. We must mention that in our case, there 
is a shift in the corresponding density peak which is caused by the curvature of the space-time, see 
Figs.~(\ref{fig:Pot_rho_SdS}) and (\ref{fig:Pot_rho_S}). Notice also that the density peak observed in 
Figs.~(\ref{fig:Pot_rho_SdS}) and (\ref{fig:Pot_rho_S}) seems to be located around the minimum of the 
induced confinement region.
The shift in the density peak caused by the geometry of the space-time, together with the position-dependent 
interaction parameter $\lambda F$ obtained above, deserves deeper regarding the stability of the 
system; this will be presented elsewhere, as it could lead to observable manifestations.

Finally, concerning the de Sitter background spacetime, we see that there is no value where the particle density 
vanishes, since the effective potential is always negative, so that a boundary is not well defined and the analogy 
evidently does not hold in this case.

%--------------------------------------------------------------------------------------------------------------------------------------------
%--------------------------------------------------------------------------------------------------------------------------------------------

\section{Conclusions and outlook}

We presented the different analogies existing between the solution to the Klein--Gordon
equation with the solution to the 
Gross--Pitaevskii equation. 

For the case of a particle trapped in a 
one dimensional box potential with a large separation between the walls (a static soliton 
kink solution), we showed that both solutions are formally related.

Additionally, we have shown that the Klein--Gordon equation in some classes of curved background spacetimes 
is such that the gravitational background induces a kind of confinement effective potential which allows quasi--stationary 
states for the scalar field, in close analogy to what happens with Bose--Einstein condensates in atomic physics. It was shown 
that a Schwarzschild--de Sitter black hole background spacetime, together with
the scalar field potential parameters, provides an effective trapping potential which allows the existence 
of such quasi--stationary
scalar field distributions. 

The curved background also induces an effective self--interaction parameter, which clearly modifies the 
strength of interactions within the system. This fact could in principle be interpreted as a kind of Feshbach 
resonance caused by the curvature of the space time. In usual Bose--Einstein condensates, the Feshbach 
resonances make it possible to tune scattering lengths and other quantities, adjusting an external field such 
as the magnetic field \cite{pethick2002}. This could affect the stability of the cloud. Thus, it is interesting to 
explore if the system is stable, taking into account the effects caused by the induced effective self--interaction 
parameter. This topic deserves deeper investigation. In particular, in order that our system
can be compared with observations, it is mandatory to derive a numerical solution, along the lines of the 
one in \cite{Claus13},
and compare with the corresponding Thomas-Fermi solution presented in this work. Indeed, the stability or
quasi-stability of the system, show in the case without self interaction, is also a question which
has to be solved, i.e. what is the influence of the $\lambda F$ term on the stability of the system?

Finally, this approach can be extended to more general scenarios, in which the spacetime has rotation, and an
analogy could be made with phenomena associated with Bose-Einstein condensates such as vorticity and superfluidity,
along the lines of \cite{Xiong14}. 
It is worth noting that these ideas could imply the possibility that a scalar field considered as 
a Bose--Einstein condensate could account for dark matter halos surrounding galaxies \cite{Barranco11, Barranco12, Barranco13}.
Clearly these topics deserve deeper investigation, which will be presented elsewhere \cite{Nuevo_nos}. 

To summarize: In this work we considered a classical relativistic scalar field as a test particle/field. For some 
particular background spacetimes, a remarkable analogy between the Klein--Gordon equation for a 
test scalar particle/field  and the Gross--Pitaevskii equation for the order parameter of a Bose--Einstein condensate trapped by an 
external potential, can be made. It is important to stress that the gravitational background provides, in 
a natural way, an effective confinement potential for the scalar field, together with an effective self--interaction parameter between the constituents of the system.

It would be desirable to have a deeper understanding of the analogy by means of a special relativistic formulation of the
Bose--Einstein condensates, and to look for some phenomena which are seen in the Bose--Einstein condensates in the laboratories. This could point to an observable signature in the astrophysical realm.

%--------------------------------------------------------------------------------------------------------------------------------------------
%--------------------------------------------------------------------------------------------------------------------------------------------
\acknowledgments

This research was supported by CONACyT Grant No. 166041F3, DGAPA--UNAM Grants No. IN115311, and 
No. IN103514. E. C. acknowledges CONACyT for the postdoctoral grant received and also CONACyT 
M\'exico, CB-2009-01, No. 132400, CB-2011, No. 166212,  and I0101/131/07
C-234/07. 
C. E-R. thank Phil Bull for his opinion on the manuscript.  
This work is part of the Instituto Avanzado de Cosmolog\'ia (IAC) collaboration (http://www.iac.edu.mx/).

%--------------------------------------------------------------------------------------------------------------------------------------------
%--------------------------------------------------------------------------------------------------------------------------------------------

\end{document}